\documentclass[journal=jacsat]{achemso}
\usepackage[version=3]{mhchem} 
\usepackage{color,soul}
\usepackage{threeparttable}
\usepackage{float}
\usepackage{adjustbox}
\usepackage{slashbox}
\usepackage{multirow}
\usepackage{booktabs}
\usepackage{times}
\usepackage{url}
\usepackage{amsmath}
\usepackage{latexsym}
\usepackage{array}
\usepackage{graphicx}
\usepackage{geometry}
\usepackage{multirow}
\usepackage{graphicx}
\usepackage{float}
\usepackage{subfigure}
\usepackage[utf8]{inputenc} 
\usepackage[font={small}]{caption} 
\usepackage{caption}

\newcommand{\erf}[1]{\operatorname{erf}\left(#1\right)}

\author{Wan-Lu Li$^{1-3}$, Kaixuan Chen$^{1-3}$, Elliot Rossomme$^{1,2}$, Martin Head-Gordon$^{1-3}$, Teresa Head-Gordon$^{1-5}$}

\affiliation{$^1$Kenneth S. Pitzer Center for Theoretical Chemistry, $^2$Department of Chemistry, $^3$Chemical Sciences Division, Lawrence Berkeley National Laboratory,
$^4$Department of Chemical and Biomolecular Engineering, $^5$Department of Bioengineering,
University of California, Berkeley,\\
Berkeley, California 94720, USA\\
}

\email{thg@berkeley.edu}
\title{Greater Transferability and Accuracy of Norm-conserving Pseudopotentials using Nonlinear Core Corrections}

\begin{document}

%
%
%
%
%
%
\clearpage
\begin{abstract}
\noindent
We present an investigation into the transferability of pseudopotentials (PPs) with a nonlinear core correction (NLCC) using the Goedecker, Teter, and Hutter (GTH) protocol across a range of pure GGA, meta-GGA and hybrid functionals, and their impact on thermochemical and non-thermochemical properties. The GTH-NLCC PP for the PBE density functional demonstrates remarkable transferability to the PBE0 and $\omega$B97X-V exchange-correlation functionals, and relative to no NLCC, improves agreement significantly for thermochemical benchmarks compared to all-electron calculations. On the other hand, the B97M-rV meta-GGA functional performs poorly with the PBE-derived GTH-NLCC PP, which is mitigated by reoptimizing the NLCC parameters for this specific functional. The findings reveal that atomization energies exhibit the greatest improvements from use of the NLCC, which thus provides an important correction needed for covalent interactions relevant to applications involving chemical reactivity. Finally we test the NLCC-GTH PPs when combined with medium-size TZV2P molecularly optimized (MOLOPT) basis sets which are typically utilized in condensed phase simulations, and show that they lead to consistently good results when compared to all-electron calculations for atomization energies, ionization potentials, barrier heights, and non-covalent interactions, but lead to somewhat larger errors for electron affinities.
\begin{figure}[H]
    \centering
\includegraphics[scale=1.0]{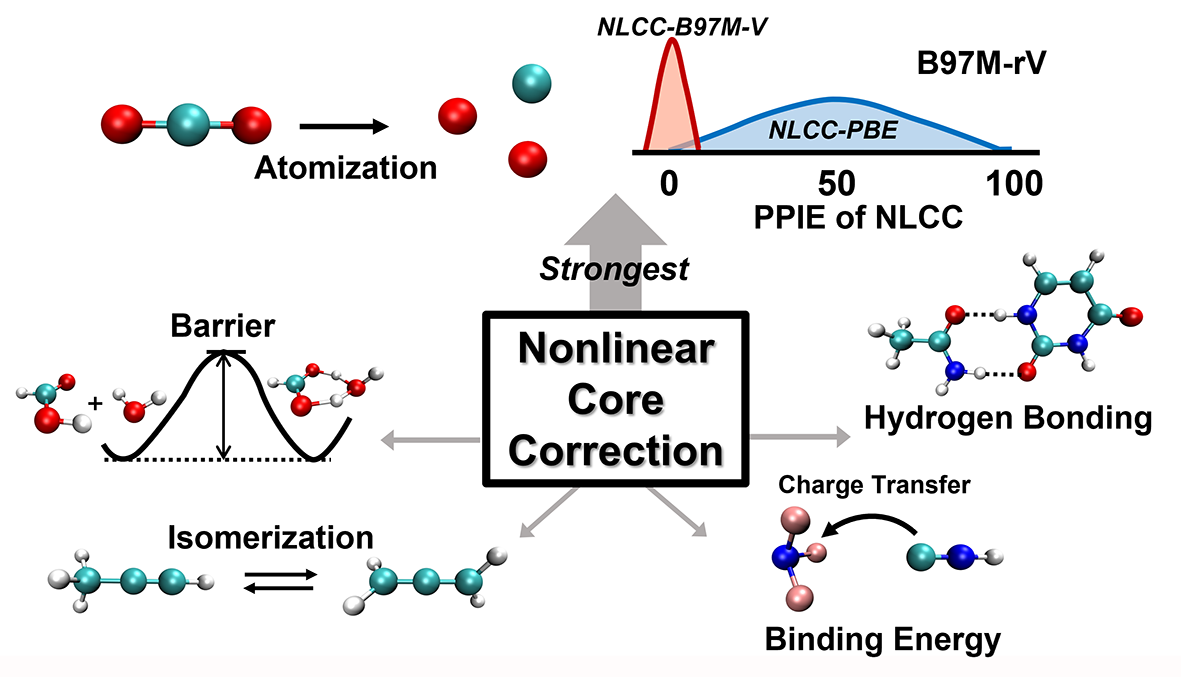}
\end{figure}
\end{abstract}

\clearpage
\vspace{15pt}
\noindent

\section{Introduction}
Pseudopotentials are mathematical representations of the interactions between explicitly treated valence electrons and effective cores that replace atomic nuclei and inner shell electrons, allowing for a substantial reduction in computational resources required to model large chemical and materials systems, while also accounting for relativistic effects in some cases.\cite{Schwerdtfeger2011} The Goedecker, Teter, and Hutter (GTH) formalism and optimization protocol has produced separable, norm-conserving, Gaussian-type dual-space pseudopotentials (PPs)\cite{PhysRevB.54.1703,PhysRevB.58.3641} that are widely used in ab initio molecular dynamics (AIMD) simulations with a mixed Gaussian-planewave (GPW) strategy\cite{doi:10.1080/002689797170220}. The GTH PP construction and  optimization is generally performed utilizing the neutral atomic state as a reference with spherical symmetry. To make an effective PP, it is essential to divide the space into two regions: muffin-tin spheres centered on the atom in a molecule or solid, and the interstitial region that encompasses the rest of the charge density\cite{martin2004electronic}. Knowing the scattering properties on the surface of the muffin-tin spheres then allows for exact solutions of the Schrödinger equation in the interstitial region.\cite{PhysRev.127.276} By satisfying the norm conservation condition in GTH PPs\cite{PhysRevB.26.4199}, the logarithmic derivative function associated with the energy accurately represents the scattering properties of a muffin-tin sphere that contains the charge distribution of the reference configuration. Screening effects give rise to an approximately invariant muffin-tin sphere, wherein the total electronic charge distribution remains largely unaffected by changes in the outer chemical environment.\cite{PhysRevA.45.88} The construction of the traditional GTH\cite{PhysRevB.54.1703} and Hartwigsen-Goedecker-
Hutter (HGH)\cite{PhysRevB.58.3641} PPs followed this approach, resulting in PPs that exhibit excellent transferability for non-spin-polarized systems. 

However, in a self-consistent field calculation, the charge distribution undergoes changes when a free atom is inserted into a molecule or solid, and the potential within the muffin-tin region will generally differ from the potential within a muffin-tin sphere of the same radius around the reference atom. As a result, the scattering properties undergo modifications, and the PP constructed based on the isolated atom might not accurately reproduce the altered scattering properties due to the new chemical environment. In spin-polarized calculations, the concept of an invariant muffin-tin sphere is also not applicable since the spin polarization deviates across different chemical environments, despite the similarity in the total charge within the invariant muffin-tin sphere. In other words, the typical assumption for PPs that assume a linear relationship between the charge distribution and potential leads to inaccuracies because the electron-electron interactions in the core region are highly localized such that they exhibit a nonlinear relationship between the electron density and the associated potential. 

One approach to address this is by incorporating nonlinear core corrections (NLCC)\cite{PhysRevB.26.1738} into the PP optimization. For the NLCC schemes, the spin and charge densities within the muffin-tin sphere are not solely determined by the valence electrons, but instead encompass the combined contributions of the valence charge and the core charge. The NLCC method allows for a more realistic treatment of the core electron behavior by taking into account the strong electron-electron interactions and the asymmetric charge density distribution near the atomic nucleus. In accordance with the principles underlying the GTH PPs, Willand et al. in 2013 put forward a NLCC method in which the core charge density is represented by a single Gaussian function\cite{doi:10.1063/1.4793260}, which captures the essential characteristics of the core charge. The amplitude and width of this Gaussian core charge distribution were subsequently optimized through a rigorous fitting procedure similar to the regular GTH PP  optimization\cite{doi:10.1021/acs.jctc.9b00553,doi:10.1021/acs.jctc.1c00026,doi:10.1021/acs.jpclett.1c02918} but including not just ground state but also excited states and ionized electronic configurations. By employing this methodology, the NLCC-PP achieves a high level of precision in describing the atomization energies in molecular systems compared to all-electron calculations and reliability for high-pressure phases of crystalline solids. Together with an adequate treatment of dispersion, such as the empirical Grimme D2\cite{https://doi.org/10.1002/jcc.20495} and D3\cite{10.1063/1.3382344} schemes, the inclusion of NLCC corrections allows for a comprehensive description of weakly bound intermolecular interactions, achieving an average error of approximately 0.5 kcal/mol.

Recently, we presented a study on the optimization and transferability of small and large core GTH PPs for Density Functional Theory (DFT) functionals applied to molecular systems and condensed phase simulations for electrocatalysis\cite{doi:10.1021/acs.jpclett.1c02918}. We also have recently conducted a systematic exploration of the extent of PP inconsistency errors (PPIEs)\cite{doi:10.1021/acs.jctc.3c00089} that arise from the transfer of a different PP relative to the chosen level of DFT when evaluating energy differences. To address these errors, we have developed and implemented empirical atom- and density functional approximation specific PP  corrections.\cite{doi:10.1021/acs.jctc.3c00089} But in neither case have we considered the role of NLCC PPs and their influence on the accuracy possible with different DFT functionals outside the original parameterization of the PBE functional\cite{doi:10.1063/1.4793260} and when combined with less complete basis sets such as the molecularly optimized (MOLOPT) basis sets used in AIMD application studies in CP2K\cite{https://doi.org/10.1002/wcms.1159}. 

In this work, we directly use the 2013 NLCC HGH-PP developed for the PBE functional\cite{doi:10.1063/1.4793260}, and investigate its transferability and accuracy across various DFT functionals including the hybrid GGA PBE0\cite{doi:10.1063/1.478522}, the semi-empirical hybrid $\omega$B97M-rV\cite{doi:10.1063/1.4952647}, and the meta-GGA B97M-rV\cite{doi:10.1021/acs.jpclett.6b02527, doi:10.1063/1.4907719}, as measured on performance for thermochemical properties, barrier heights, isomerization energies, and non-covalent interactions. In addition, we also combine the GTH-NLCC method with the medium-sized TZVP MOLOPT basis sets that are used to enhance the computational efficiency in CP2K\cite{doi:10.1063/1.2770708} and compare them to the more complete def2-TZVPPD basis set results. Although the NLCC GTH-PP results are more transferable and yield excellent results for the hybrid functionals regardless of basis set, the B97M-rV functional exhibits a large PPIE. Hence we undertook a reoptimization process to enhance the accuracy of the NLCC parameters for this meta-GGA functional which brings it into line with the other DFT/basis set combinations. As we show below, the use of the NLCC correction improves overall accuracy over the full range of DFT functionals and basis sets considered here, minimizing errors across all data sets but especially for thermochemical properties that are most relevant for applications involving chemical reactivity.

\section{Computational Details}
\paragraph{GTH PPs and Optimization Algorithm of NLCC Parameters.} The norm-conserving, dual-space, and explicitly separable GTH pseudopotential $V^{pp}$ contains local and nonlocal parts,\cite{PhysRevB.54.1703,Krack2005} where the local term is expressed as 
\begin{equation}
  \text{V}^{pp}_{loc}(r) = -\dfrac{\text{Z}_\text{ion}}{r}
  \erf{\dfrac{r}{\sqrt{2}\text{r}^\text{pp}_\text{loc}}}
  +\sum_{i}^{4}\text{C}^\text{pp}_{i}
  \left(
    \dfrac{r}{\text{r}^\text{pp}_\text{loc}}
  \right)^{2i-2}
  \exp\left[
    -\dfrac{1}{2}
    \left(
      \dfrac{r}{\text{r}^\text{pp}_\text{loc}}
    \right)^2
  \right] 
  \label{eqn:1}
\end{equation}
and erf represents the error function, $\text{Z}_{ion}$ is the ionic charge of the atom,  $\text{r}^\text{pp}_\text{loc}$ controls the range of the Gaussian ionic charge distribution, and the $\text{C}^\text{pp}_{i}$ are the coefficients to be optimized. 

The nonlocal part is expressed as
\begin{equation}
  \text{V}^\text{pp}_{nl}(\boldsymbol{r},\boldsymbol{r}^\prime) = 
  \sum_{lm}\sum_{ij}\left \langle \boldsymbol{r}|\text{p}^{lm}_{i} \right \rangle\text{h}^{l}_{ij}\left \langle \text{p}^{lm}_{j}|\boldsymbol{r^\prime} \right \rangle  \label{eqn:2}
\end{equation}
where $\text{h}^{l}_{ij}$ denotes a scattering matrix element to be fitted and the Gaussian-type projectors are given as
\begin{equation}
  \left \langle \boldsymbol{r}|\text{p}^{lm}_{i} \right \rangle = \text{N}^{l}_{i}\text{Y}^{lm}(r)\text{r}^{l+2i-2}
  \exp\left[
    -\dfrac{1}{2}
    \left(
      \dfrac{r}{r_l}
    \right)^2
  \right] \label{eqn:3}
\end{equation}
where \textit{l} denotes the angular momentum quantum number of subshells, $\text{N}^{l}_{i}$ denotes a normalization constant and $\text{Y}^{lm}$ is a spherical harmonic function. Therefore, parameters r$_{loc}$ and C$_i$ in the local term, and h$_{ij}$ and r$_l$ in the nonlocal part need to be fitted in the GTH PP optimization. 

The nonlinear core correction is given by 

\begin{equation}
    \rho_{c}(r)=c_{core}\frac{Z-Z_{ion}}{(\sqrt{2\pi}r_{core})^3}e^{-\frac{r^{2}}{2r_{core}^2}} \label{eqn:4}
\end{equation}

\noindent
in which the core charge is described by a single Gaussian function, which is efficient for numerical integration, and is then incorporated into the generalized Kohn-Sham total energy equation:

\begin{multline}
    E_{KS}=\sum_{i}\langle{\psi}_{i}|\lbrace-\frac{1}{2}[\nabla^{2}+\nabla\cdot(V_{\tau}[\rho](\mathbf{r})\nabla)]+V_{H}[\rho]+V_{xc}[\rho+\rho_{c}]+V^{pp}\rbrace|\psi_{i}\rangle\\-E_{H}[\rho]+E_{xc}[\rho+\rho_{c}]-\int d\mathbf{r}\rho(\mathbf{r})V_{xc}[\rho+\rho_{c}](\mathbf{r}) \label{eqn:5}
\end{multline}

\noindent
where $E_{xc}$, $V_{xc}$ and $V_{\tau}$ are the exchange-correlation energy, potential, and kinetic energy term, respectively, and $V_{H}$ is the Hartree potential. In the CP2K code\cite{https://doi.org/10.1002/wcms.1159}, it is assumed that the kinetic energy density originating from the core orbitals is not included. Therefore, when performing meta-GGA calculations, the variables used are the core charge augmented density and density gradient, along with the kinetic energy density solely from the valence orbitals. As a result, in our NLCC optimization for the meta-GGA functional, we develop a model for the core density and its gradient, adding it to the valence density. However, it is important to note that the correction for the kinetic energy density is still missing due to this assumption.

The amplitude ($c_{core}$) and width ($r_{core}$) of the function are optimized based on the usual procedure of modulating initial step size, weight for configurations in different regions such as core, valence and virtual states, and target accuracy for Kohn-Sham eigenvalues, with reference to the all-electron results with the consideration of scalar relativistic effects.\cite{PhysRevB.58.3641} However, contrary to the optimization procedure of a traditional GTH PP, which is fitted to a single atomic configuration considered as a symmetric sphere, the optimized NLCC PPs are parameterized not only with the ground state, but also with  excited states and ionized electronic configurations. Several atomic properties were chosen in the fitting procedure with respect to the all-electron (AE) calculations: 1) Atomic eigenvalues of the occupied and first few unoccupied valence orbitals, 2) Charge within the inert region ($r_{loc}$) of the pseudo atom matching the charge in the same region for all the orbitals; this criterion illustrates that the PP is norm conserving for all the configurations used in the optimization, 3) Total energy difference, and 4) Spin polarization energy of all-electron calculation is reproduced. This is incorporated into the objective penalty function expressed in Eq.~\ref{eqn:6}
\begin{equation}
  S = \sum_{n,l}\text{w}^{2}_{n,l}(\text{p}^\text{AE}_{n,l}-\text{p}^\text{PP}_{n,l})  \label{eqn:6}
\end{equation}
where $p$ denotes different atomic properties described above, $w$ is the weight of each property, and $n$ and $l$ are principal and angular momentum quantum numbers, respectively. 

\textbf{Computational and Theoretical Details.} The ATOM module implemented in CP2K package\cite{https://doi.org/10.1002/wcms.1159} was used to optimize the new NLCC GTH PPs at the level of B97M-V. The NLCC terms optimized on top of Hartwigsen-Goedecker-Hutter (HGH) pseudopotential served as an initial guess \cite{doi:10.1063/1.4793260}. In the calculations for different applications of the new PPs, the energy cutoff was set as 1200 Ry with a box size of 20 \AA $ \times$ 20 \AA $ \times$ 20 \AA. During the Self-Consistent-Field (SCF) calculation, a strategy of orbital transform (OT)\cite{doi:10.1063/1.1543154,doi:10.1063/1.2841077} is utilized to accelerate the convergence, of which the threshold is set to be $10^{-6}$ hartree. The reference results are obtained from all-electron calculations with Gaussian-type def2-TZVPPD basis sets\cite{B508541A,doi:10.1063/1.3484283,doi:10.1063/1.1627293}.

In order to examine the NLCC correction for different chemical properties, we chose a number of benchmark data sets. We follow the convention outlined by Mardirossian and Head-Gordon\cite{doi:10.1080/00268976.2017.1333644} to organize datasets into categories: (1) thermochemistry (TC), (2) barrier heights (BH), (3) isomerization energies (I), and (4) non-covalent interactions (NC), ranging from most difficult to easier in regards demands on the PP accuracy. Note however we are not using their MGDB84 dataset here. For the TC category we used the whole series of G2 datasets\cite{doi:10.1063/1.473182,doi:10.1063/1.456415,doi:10.1063/1.458892} that evaluates the covalent bond formation energy, as well as the TAE140nonMR and TAE140MR\cite{KARTON2011165} for easy and difficult atomization energies, respectively. For the BH category we used HTBH38\cite{B416937A}, NHTBH38\cite{doi:10.1021/jp045141s}, and  WCPT27\cite{doi:10.1021/jp301499y} datasets representing hydrogen transfer barrier heights, non-hydrogen transfer barrier heights, and barrier heights of water-catalysed proton-transfer reactions, respectively. For the I category we used ID (isomerization energies "difficult") and ISOMERIZATION20\cite{KARTON2011165} for isomerization energies; G21EA\cite{doi:10.1021/ct900489g,doi:10.1063/1.460205} and G21IP\cite{doi:10.1021/ct900489g,doi:10.1063/1.460205} for adiabatic electron affinities and ionization potentials of atoms and small molecules, respectively. For the NC category we used the NCED (non-covalent "easy" dimers) and S66\cite{doi:10.1021/ct200523a,doi:10.1021/jp5098603} describing binding energies between organic molecules and biomolecules, and a NCD (non-covalent "difficult") CT20\cite{doi:10.1021/ct200930x} dataset for binding energies of charge-transfer complexes.

\section{Results}
The current standard in condensed phase simulations for codes such as CP2K is to use GTH PPs that are combined with compact MOLOPT basis sets, such as  Table 1 and Figure 1(a-c) shows the resulting errors made on atomization energies with GTH-PP/MOLOPT relative to def2-TZVPPD all-electron calculations, which is the most representative combination of PPs and basis sets used in AIMD applications where chemical transformations are critical. The standard PBE PP and its transferability to the two hybrid functionals for the GTH/MOLOPT combination is overall superior

\begin{table}[]
\begin{center}
\captionsetup{font=small}
\caption{\textit{Errors in atomization energies predicted by GTH PPs compared to def2-TZVPPD all-electron calculations with or without NLCC corrections and using MOLOPT or def2-TZVPPD basis sets.} Mean Absolute Deviations (MADS, in kcal/mol) of the different DFT functionals and basis set combinations analyzed over the G2 series\cite{doi:10.1063/1.473182,doi:10.1063/1.456415,doi:10.1063/1.458892} and the TAE140 datasets\cite{KARTON2011165} divided into 124 "easy" molecules without multi-configurational character and 16 multi-reference atomization energies.}
\setlength{\tabcolsep}{1.8mm}{
\small{
\begin{tabular}{llllll}
\toprule[1.5pt]
PP &   &   GTH & GTH & GTH-NLCC & GTH-NLCC \\ \hline
Basis Set &  Dataset &   MOLOPT & def2-TZVPPD & MOLOPT & def2-TZVPPD \\ \hline
\multirow{3}{*}{PBE} & G2 & 10.13 & 15.55 & 6.06 & 1.90 \\
 & TAE140nonMR & 10.20 & 12.25 & 3.36 & 1.32 \\
 & TAE140MR & 12.61 & 9.92 & 2.61 & 1.70 \\ \hline
\multirow{3}{*}{PBE0} & G2 & 6.68 & 11.05 & 1.80 & 5.68 \\
 & TAE140nonMR & 7.69 & 8.81 & 2.34 & 3.53 \\
 & TAE140MR & 10.65 & 6.96 & 3.11 & 2.76 \\ \hline
\multirow{3}{*}{$\omega$B97X-V} & G2 & 4.98 & 9.66 & 4.03 & 3.99 \\
 & TAE140nonMR & 4.05 & 4.16 & 4.48 & 2.60 \\
 & TAE140MR & 8.25 & 4.35 & 4.26 & 4.82 \\
 \bottomrule[1.5pt]
\end{tabular}}}
\end{center}
\end{table}

\begin{figure}[H]%
\centering
\includegraphics[width=0.99\textwidth]{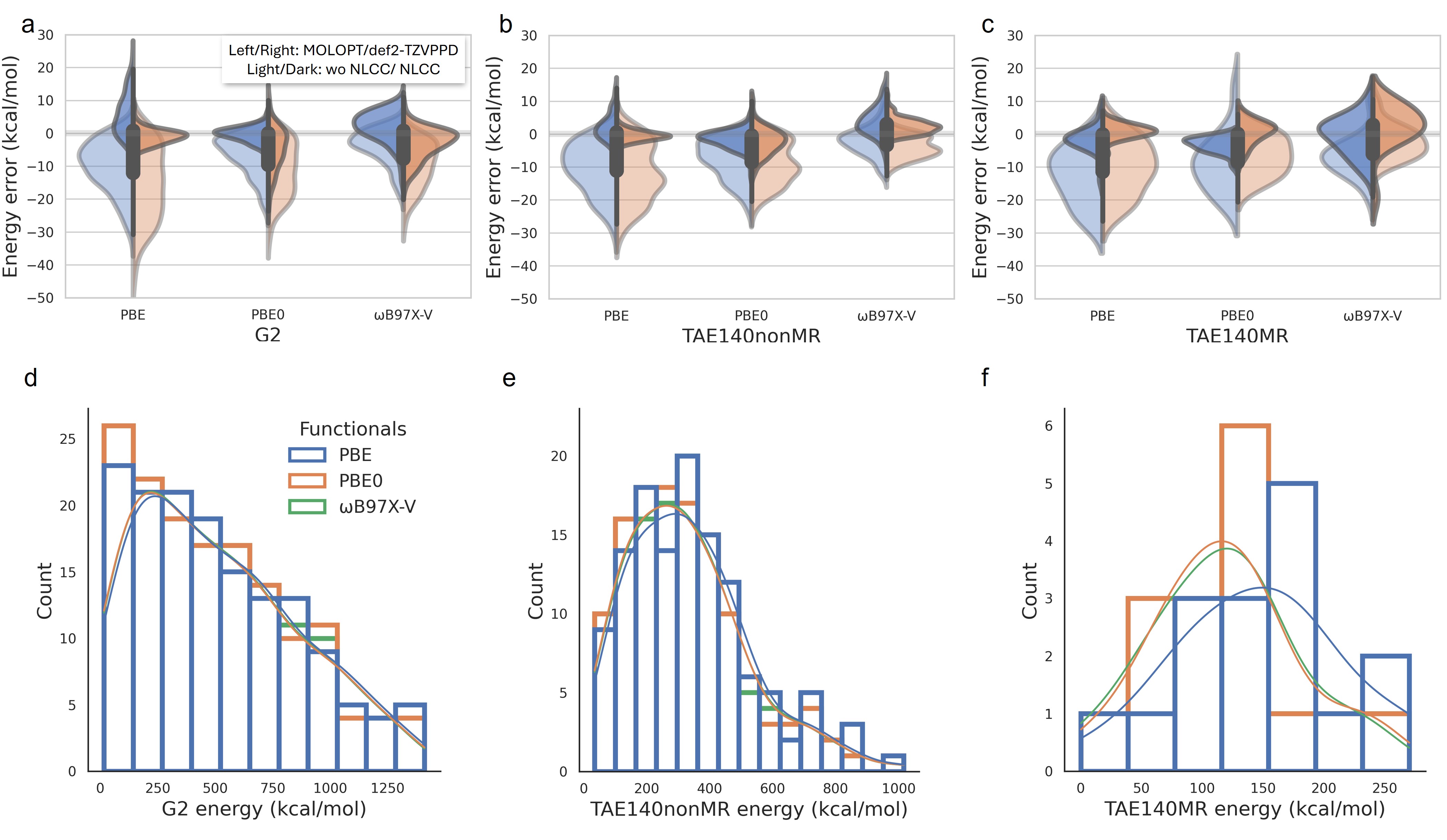}
\caption{\textit{Nonlinear core corrections and basis set effects on atomization energies for the G2 and TAE140 datasets.} We decomposed the total TAE140 dataset into 124 TAE140nonMR and 16 TAE140MR subsets. All the GTH pseudopotentials applied here are at PBE level. (a-c): violin diagrams of the energy error compared with all-electron def2-TZVPPD results for density functionals of PBE, PBE0 and $\omega$B97X-V applied to G2, TAE140nonMR and TAE140MR data set, respectively. (d-f): distribution of the all-electron results predicted for different functionals for G2, TAE140nonMR and TAE140MR datasets, respectively.}
\end{figure}

\noindent
to the larger basis set for molecules without multi-reference character, but mostly mean absolute deviations (MADs) are surprisingly large, i.e. 5-10\% in the mean absolute percentage error (MAPE) normalized by the corresponding average value of each dataset, assessed by the energy scales shown Figure 1(d-f).

To assess the errors incurred for the GTH-NLCC parameters optimized for PBE\cite{doi:10.1063/1.4793260}, Table 1 and Figure 1(a-c) shows a significant reduction in the MADs and MAPE regardless of basis set relative to the all-electron calculations. This correction effectively addresses the issue of underestimating the covalent interactions arising from the original GTH/MOLOPT parameterization. For the PBE functional, the GTH-NLCC combined with more complete basis set def2-TZVPPD yields the best results, with the MAD generally less than 2 kcal/mol and MAPE less than 1.2\%. Such consistency with the all-electron calculation has been corroborated in the study conducted by Willand et al.\cite{doi:10.1063/1.4793260} where they use the standard G2-1 dataset containing only 54 molecules. However the results using GTH-NLCC/MOLOPT, the basis set relied upon in condensed phase AIMD simulation work, is an impressive reduction in overall error as well. 

Often PPs developed for one DFT functional, typically PBE, are transferred for use with other DFT functionals. Table 1 and Figure 1(a-c) show that the underestimation error incurred by PBE can be alleviated by utilizing higher hierarchical density functionals, such as hybrid PBE0 and $\omega$B97X-V functionals that incorporate a more accurate exchange-correlation term. And yet the hybrid functionals also benefit from the NLCC PPs transferred from PBE. Notably, the combination of GTH-NLCC and MOLOPT basis sets exhibits a remarkable capacity to faithfully reproduce all-electron outcomes, when compared to the larger basis sets used in electronic structure codes. This underscores the extensive utility of MOLOPT basis in practical computations, as it adeptly strikes a fine balance between precision and computational efficiency.


Table 2 and Figure 2 depict the errors in atomization energies predicted by various combinations of PPs and basis sets when evaluated across the G2, TAE140nonMR, and TAE140MR datasets for the semi-empirical meta-GGA B97M-rV functional. Although the GTH-NLCC PP for PBE was highly transferable to the two hybrid functionals, one of which is also a semi-empirical 

\begin{table}[]
\begin{center}
\captionsetup{font=small}
\caption{\textit{Errors in atomization energies predicted by GTH PPs compared to def2-TZVPPD all-electron calculations with or without NLCC corrections and using MOLOPT or def2-TZVPPD basis sets for the meta-GGa B97M-rV functional.} Mean Absolute Deviations (MADS, in kcal/mol) of the different DFT functionals and basis set combinations analyzed over the G2 series\cite{doi:10.1063/1.473182,doi:10.1063/1.456415,doi:10.1063/1.458892} and the TAE140 datasets\cite{KARTON2011165} divided into 
the 124 "easy" molecules without multi-configurational character and 16 multi-reference atomization energies.}
\setlength{\tabcolsep}{0.5mm}{
\small{
\begin{tabular}{lllllll}
\toprule[1.5pt]
PP & GTH & GTH & GTH-NLCC & GTH-NLCC & GTH-NLCC-OPT & GTH-NLCC-OPT\\ \hline
Basis Set & MOLOPT & def2-TZVPPD & MOLOPT & def2-TZVPPD & MOLOPT & def2-TZVPPD\\ \hline
Dataset & & & &  &  & \\ \hline
G2 & 25.42 & 29.96 & 68.76 & 61.39 & 6.63 & 5.95 \\
TAE140nonMR & 19.38 & 20.36 & 55.23 & 51.93 & 5.62 & 4.16 \\
TAE140MR & 11.87 & 9.07 & 38.71 & 42.43 & 6.25 & 5.24 \\
 \bottomrule[1.5pt]
\end{tabular}}}
\end{center}
\end{table}

\begin{figure}[H]%
\centering
\includegraphics[width=0.99\textwidth]{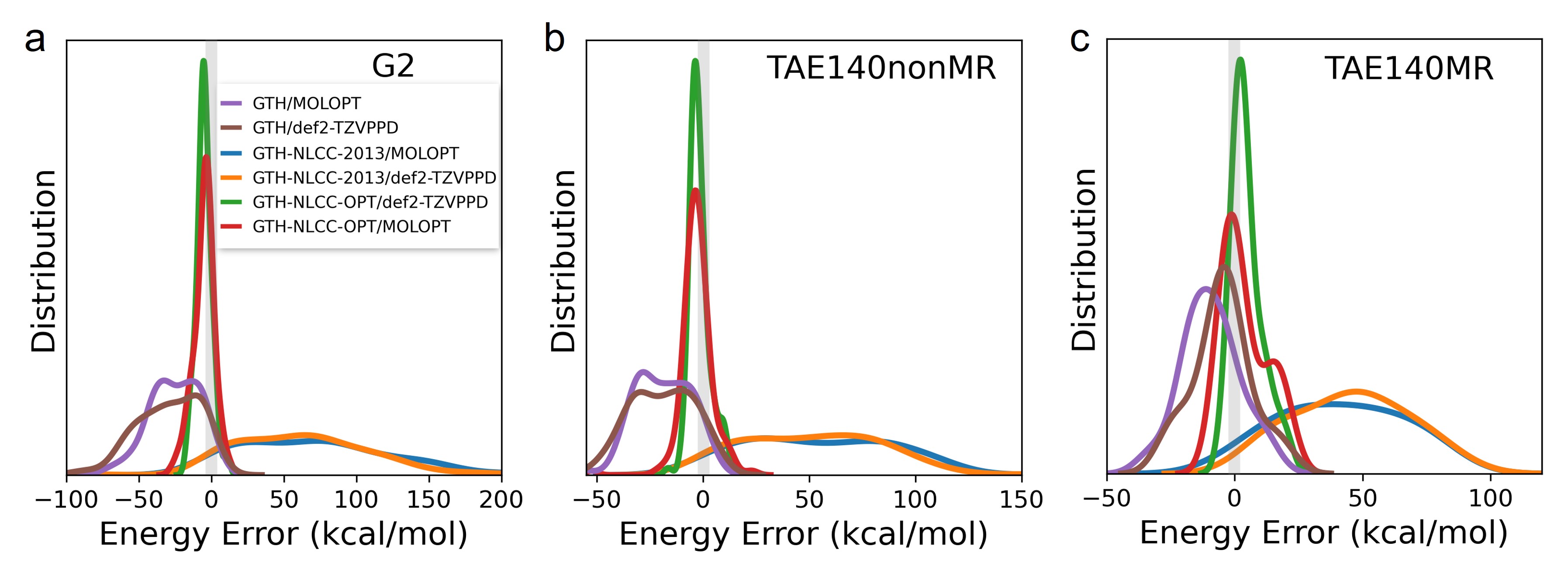}
\caption{\textit{Energy error distribution of atomization energies for the G2 and TAE140 datasets predicted by B97M-rV functional.} The reference is all-electron calculation with def2-TZVPPD basis sets.}
\end{figure}

\noindent
functional developed by the same data-driven procedure\cite{C3CP54374A}, we find the transferability is quite poor for B97M-rV meta-GGA functional\cite{doi:10.1021/acs.jpclett.6b02527,doi:10.1063/1.4907719}, yielding large MAD errors ($>$ 50 kcal/mol) regardless of the completeness of the basis sets when compared to all-electron calculations (Table 2). Notably two clear trends emerge, demonstrating first that the original GTH/MOLOPT approach significantly underestimates the atomically covalent interaction, whereas the existing NLCC correction (GTH-NLCC-2013) exhibits an adverse and now substantial overestimation of the interaction. By comparing the MADs predicted by the original GTH/MOLOPT approach (Table S2), it is also apparent that under the existing NLCC correction the lighter elements exhibit significantly larger MADs compared to their heavier counterparts. This disparity is likely attributed to the greater sensitivity to the core region for lighter elements, while conversely the stronger screening effect\cite{PhysRevA.45.88} exhibited by heavier elements helps maintain the invariance of the charge density in the core region, resulting in lower MAD values. Nonetheless, we found it imperative in this case to reparameterize the NLCC PPs specifically for the B97M-rV functional. After reoptimization of the NLCC parameters at the level of B97M-rV for only the lighter main-group elements (B, C, N, O, F), leaving the original GTH parameters in place for the heavier atoms, we get more consistent results with a MAD less than 6 kcal/mol using either the MOLOPT or more complete basis sets (Table 2 and Figure 2). 

\begin{figure}[H]%
\centering
\includegraphics[width=0.99\textwidth]{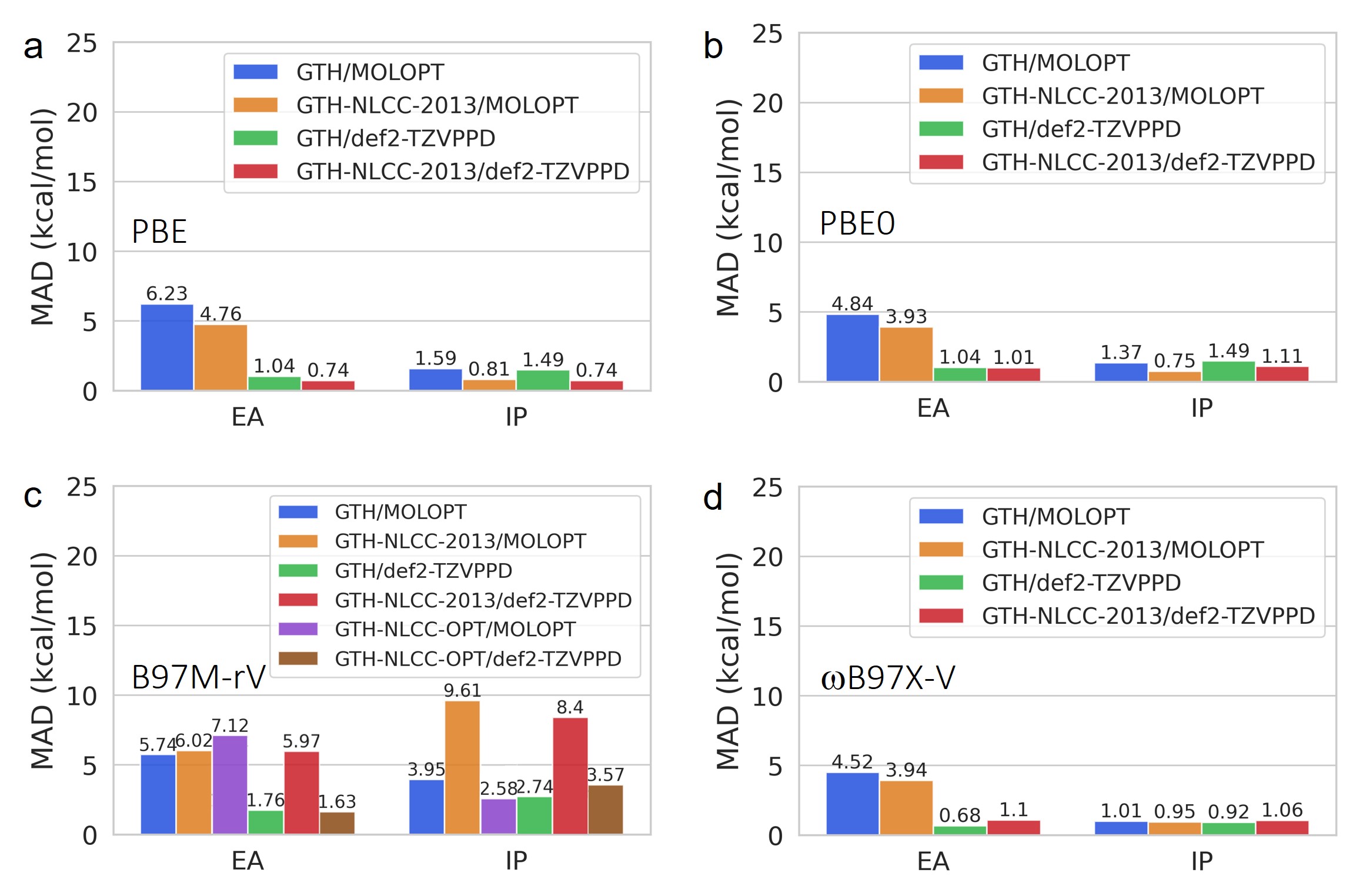}
\caption{\textit{Sensitivity of electron affinity (EA) and ionization potential (IP) to basis size and NLCC effect.} In Figure 3c from B97M-rV method, the additional bars are corresponding to the results predicted by optimized NLCC parameters for lighter elements (B, C, N, O, F) and original GTH for heavier elements (Al, Si, P, S, Cl).}
\end{figure}

Additionally, we investigated the transferability and errors for the standard and NLCC PPs and basis set combinations for other thermochemical properties including electron affinities (EA) and ionization potentials (IP) derived from the G21EA and G21IP datasets\cite{doi:10.1063/1.460205,doi:10.1021/ct100466k} as shown in Figure 3. As detailed in Figures S1 and S2, ionization involves a larger energy scale than that found for electron affinity energies, as it is harder to remove an electron from the highest occupied molecular orbital (HOMO) than to attach an electron into the lowest unoccupied molecular orbital (LUMO) because the former is much deeper than the latter relative to the vacuum level. Overall we find that IPs are fairly insensitive to both PPs and basis set completeness for the pure GGA and hybrid DFT functionals, whereas the B97M-rV meta GGA benefits from NLCC optimization, after which it yields comparable results to the other DFT functionals and regardless of basis set. These figures also highlight a significant trend in which the EA is far more sensitive to the completeness of the basis set, and that the NLCC has a more minor corrective effect on the errors due to overbinding. 

\vspace{5mm}
\begin{table}[]
\captionsetup{font=small}
\caption{\textit{Errors in non-thermal chemistry properties predicted by GTH PPs compared to def2-TZVPPD all-electron calculations with or without NLCC corrections and using TZV2P MOLOPT or def2-TZVPPD basis sets.} MAD results (kcal/mol)  predicted by PBE, PBE0 and $\omega$B97M-rV functionals for S66, CT20, ISOMERIZATION20, HTBH38, NHTBH38, and WCPT27.} 
\vspace{-20pt}
\begin{center}
\setlength{\tabcolsep}{1mm}{
\small{
\begin{tabular}{llllllll}
\toprule[1.5pt]
Functional & PP/Basis & S66 & CT20 & ISOMER20 & HTBH38 & NHTBH38 & WCPT27 \\ \hline
\multirow{4}{*}{PBE} & GTH/MOLOPT & 0.12 & 0.11 & 0.63 & 1.13 & 2.00 & 0.43 \\
 & GTH/def2-TZVPPD & 0.06 & 0.04 & 0.99 & 0.95 & 1.00 & 0.90 \\
 & GTH-NLCC/MOLOPT & 0.14 & 0.10 & 1.57 & 0.38 & 1.84 & 0.49 \\
 & GTH-NLCC/def2-TZVPPD & 0.08 & 0.05 & 0.72 & 0.61 & 0.88 & 1.15 \\ \hline
\multirow{4}{*}{PBE0} & GTH/MOLOPT & 0.22 & 0.12 & 0.74 & 0.83 & 1.67 & 0.44 \\
 & GTH/def2-TZVPPD & 0.07 & 0.07 & 0.78 & 0.71 & 1.03 & 1.04 \\
 & GTH-NLCC/MOLOPT & 0.23 & 0.12 & 0.68 & 0.37 & 1.87 & 0.45 \\
 & GTH-NLCC/def2-TZVPPD & 0.12 & 0.08 & 0.78 & 0.76 & 1.47 & 1.31 \\ \hline
\multirow{4}{*}{$\omega$B97M-rV} & GTH/MOLOPT & 0.25 & 0.11 & 0.89 & 0.52 & 1.41 & 0.62 \\
 & GTH/def2-TZVPPD & 0.10 & 0.09 & 0.38 & 0.51 & 0.58 & 1.03 \\
 & GTH-NLCC/MOLOPT & 0.24 & 0.11 & 0.93 & 0.23 & 1.63 & 0.57 \\
 & GTH-NLCC/def2-TZVPPD & 0.11 & 0.08 & 0.37 & 0.51 & 1.03 & 1.28 \\ \hline
\multirow{4}{*}{B97M-rV} & GTH/MOLOPT & 0.10 & 0.14 & 1.56 & 1.36 & 2.19 & 0.50 \\
& GTH/def2-TZVPPD & 0.22 & 0.09 & 1.46 & 1.20 & 1.53 & 0.80 \\
& GTH-NLCC-2013/MOLOPT & 0.30 & 0.21 & 5.74 & 3.96 & 2.40 & 2.05 \\
& GTH-NLCC-2013/def2-TZVPPD & 0.24 & 0.17 & 4.67 & 4.00 & 2.06 & 2.70 \\ 
\bottomrule[1.5pt]
\multirow{2}{*}{B97M-rV} & GTH-NLCC-OPT/MOLOPT & 0.27 & 0.16 & 0.61 & 0.79 & 2.58 & 0.62 \\
& GTH-NLCC-OPT/def2-TZVPPD & 0.23 & 0.11 & 0.42 & 0.67 & 0.75 & 0.57 \\ \bottomrule[1.5pt]
\end{tabular}%
}}
\end{center}
\end{table}

Table 3 provides an assessment of PP/basis set combinations for other non-reactive  properties including binding energies of non-covalent interactions in organic molecules and biomolecules (S66)\cite{doi:10.1021/ct200523a,doi:10.1021/jp5098603}, binding energies of charge-transfer complexes (CT20)\cite{doi:10.1021/ct200930x}, isomerization energies (ISOMERIZATION20)\cite{KARTON2011165}, hydrogen transfer barrier heights (HTBH38)\cite{B416937A}, non-hydrogen transfer barrier heights (NHTBH38)\cite{doi:10.1021/jp045141s}, and barrier heights for water-catalyzed proton-transfer reactions (WCPT27)\cite{doi:10.1021/jp301499y}. In Table 3 we consider the accuracy and transferability of the PBE PP to the other functionals. Clearly, the NLCC effect is negligible for these non-thermochemical properties as the chemical environment is not dramatically changed before and after the reaction and the spin polarized state is less involved compared to the atomization process. Similar to the other properties examined previously, the GTH-NLCC pseudopotential at the PBE level exhibits poor transferability to the B97M-rV functional, resulting in significantly worse results compared to the GTH/MOLOPT approach without correction. However, with utilization of optimized NLCC parameters for the B97M-rV functional, substantial improvements can be achieved for non-thermal properties as well.

\section{Discussion and Conclusion}
Pseudopotentials are widely employed in theoretical chemistry to provide a smooth potential profile within a specified core region, that permits removal of a specified set of core electrons from explicit consideration, thereby reducing the computational overhead and often enabling efficient inclusion of relativistic effects. However, their utilization can introduce non-negligible deviations, particularly in spin-polarized systems, as spin polarization exhibits distinct variations in response to different chemical environments, despite the apparent similarity in charge density within the invariant muffin-tin sphere. To overcome this challenge, the NLCC strategy developed for HGH pseudopotentials reported by Willand et al. \cite{doi:10.1063/1.4793260} introduced a single Gaussian function as the core charge density, thereby attaining chemical accuracy when compared to all-electron calculations for the standard G2-1 dataset for the PBE functional as well as for semiempirical models corrected by Grimme dispersion terms. 

In this work we tested the transferability and accuracy of the original NLCC parameters to the GTH pseudopotential developed for the PBE GGA functional\cite{PhysRevLett.77.3865} and associated with the TZV2P MOLOPT basis set\cite{doi:10.1063/1.2770708}, to hybrid DFT functionals including  PBE0\cite{doi:10.1063/1.478522} and the range-seperated hybrid, meta-GGA $\omega$B9X-V\cite{doi:10.1063/1.4952647}, as well as the B97M-rV meta-GGA functional. Our investigation of transferability and accuracy encompassed a wide range of both thermochemistry and non-thermochemistry datasets, and our findings indicate that the NLCC correction has the largest impact on atomization energies, effectively rectifying the original GTH/MOLOPT method and resulting in an error reduction of less than 1\% when compared to all-electron calculations. 

For all properties investigated, the existing NLCC parameters optimized at the level of PBE have good transferability to PBE0 and $\omega$B97M-rV while maintaining satisfactory accuracy using the TZV2P MOLOPT basis sets, although electron affinities were found to be more sensitive to the completeness of basis sets than the NLCC correction. In contrast, the GTH-NLCC PP for PBE shows poor transferability to the meta-GGA B97M-rV functional, with significant deviations of over 50 kcal/mol in atomization energies when compared to all-electron calculations, thus requiring a reoptimization of the NLCC parameters specifically for this DFT functional. We note that similar PPIEs were found for the SCAN functional as described by Rossomme and co-workers\cite{doi:10.1021/acs.jctc.3c00089}, indicating that meta-GGA functionals may be unusually sensitive to PP replacement of core electrons. Fortunately, our optimization of the GTH-NLCC PP for B97M-rV overcame the drastic errors using the standard GTH/MOLOPT, in which optimization was only necessary for lighter second-row elements while heavier elements did not require any changes, which may be insightful for improving other meta-GGA functionals. We note that the errors in the PPs for the meta-GGA may arise from the assumption made regarding the kinetic energy density, which neglects the contribution of the core charge and focuses solely on the valence density (which is also an assumption in the CP2K program). This highlights the need, and opportunity, for developing an NLCC correction for the kinetic energy density since meta-GGAs may offer better DFT accuracy at an affordable cost.

In addition to transferability, the GTH-NLCC PPs combined with the MOLOPT basis sets performed overall as well as the complete basis set calculations for PBE and hybrid functionals, and comparably for B97M-rV after optimization of its non-transferable GTH-NLCC PP. Thus our conclusion is that the GTH-NLCC PP/MOLOPT combination can be reliably employed in large-scale systems, effectively reducing computational costs while maintaining good accuracy. In summary, we believe these findings contribute to a better understanding of the sources of error in calculations using PPs and provide a protocol for enhancing the  reliability of other DFT functionals when combined with chosen PPs and basis sets.


\begin{suppinfo}

Supporting Information: Electronic configurations of core and valence regions for pseudopotentials of elements studied; Comparison of mean absolute deviations for a specific atom type within
the G2 dataset; Distributions of MADs and all-electron calculations for IP and EA datasets; NLCC performance on non-thermochemical properties.

\end{suppinfo}

\begin{acknowledgement}
\noindent 
This work was supported by the U.S. Department of Energy, Office of Science, Office of Advanced Scientific Computing, and Office of Basic Energy Sciences, via the Scientific Discovery through Advanced Computing (SciDAC) program. This work used computational resources provided
by the National Energy Research Scientific Computing Center (NERSC), a U.S. Department of Energy Office of Science User Facility operated under Contract DE-AC02-05CH11231. W.-L.L. and T.H.-G. acknowledge discussions with Prof. Jürg Hutter (University of Zurich) when optimizing the NLCC parameters.
\end{acknowledgement}

\bibliography{PP}

\end{document}


\maketitle

\clearpage
\noindent{Table S1. Electronic configurations of core and valence regions for pseudopotentials of elements studied.}
\setlength{\tabcolsep}{8mm}{
\begin{center}
\begin{tabular}{lllll}
\toprule[1.5pt]
Element & Functional & NLCC & Core & Valence \\ \hline
\multirow{4}{*}{H, Li, Be} & PBE & N & None & All \\
 & PBE0 & N & None & All \\
 & $\omega$B97M-rV & N & None & All \\
 & B97M-rV & N & None & All \\ \hline
\multirow{4}{*}{B, C, N, O, F} & PBE & Y & [He] & 2s$^2$2p$^x$ \\
 & PBE0 & Y & [He] & 2s$^2$2p$^x$ \\
 & $\omega$B97M-rV & Y & [He] & 2s$^2$2p$^x$ \\
 & B97M-rV & Y & [He] & 2s$^2$2p$^x$ \\ \hline
\multirow{4}{*}{Na, Mg} & PBE & N & [Ne] & 3s$^2$3p$^x$ \\
 & PBE0 & N & [Ne] & 3s$^2$3p$^x$ \\
 & $\omega$B97M-rV & N & [Ne] & 3s$^2$3p$^x$ \\
 & B97M-rV$^a$ & N & [Ne] & 3s$^2$3p$^x$ \\ \hline
\multirow{4}{*}{Al, Si, P, S, Cl} & PBE & Y & [Ne] & 3s$^2$3p$^x$ \\
 & PBE0 & Y & [Ne] & 3s$^2$3p$^x$ \\
 & $\omega$B97M-rV & Y & [Ne] & 3s$^2$3p$^x$ \\
 & B97M-rV$^a$ & N & [Ne] & 3s$^2$3p$^x$ \\ 
\bottomrule[1.5pt]
\end{tabular}%
\end{center}}
\tabnote{$^{\rm a}$ At the B97M-rV level, we initially tried to optimize the third-row elements. However, due to substantial errors, we were unable to obtain reliable results. Therefore, we resorted to using the GTH/MOLOPT method without B97M-rV to generate accurate results for these heavier elements.}
\clearpage

\clearpage
\noindent{Table S2. Comparison of mean absolute deviations (kcal/mol) for a specific atom type within the G2 dataset\cite{doi:10.1063/1.473182,doi:10.1063/1.456415,doi:10.1063/1.458892}. Predictions were made at the PBE and B97M-rV levels using the complete basis set def2-TZVPPD, with and without NLCC correction.}
\setlength{\tabcolsep}{2.3mm}{
\begin{center}
\begin{tabular}{lllllll}
\toprule[1.5pt]
Functional & Pseudopotential/Basis & B & C & N & O & F \\ \hline
\multirow{2}{*}{PBE} & GTH/def2-TZVPPD & 12.76 & 20.52 & 22.19 & 19.51 & 16.58 \\
 & GTH-NLCC-2013/def2-TZVPPD & 2.44 & 2.19 & 1.56 & 1.28 & 9.13 \\ \hline
\multirow{3}{*}{B97M-rV} & GTH/def2-TZVPPD & 29.90 & 29.42 & 38.98 & 24.47 & 18.25 \\
 & GTH-NLCC-OPT-all$^a$/def2-TZVPPD & 27.55 & 11.28 & 10.79 & 11.14 & 9.00 \\
 & GTH-NLCC-OPT$^b$/def2-TZVPPD & 4.07 & 6.43 & 10.70 & 6.68 & 8.06 \\ \hline
Functional & Pseudopotential/Basis & Al & Si & P & S & Cl \\ \hline
\multirow{2}{*}{PBE} & GTH/def2-TZVPPD & 4.11 & 2.83 & 2.91 & 9.40 & 7.71 \\
 & GTH-NLCC-2013/def2-TZVPPD & 8.34 & 2.23 & 1.30 & 1.21 & 6.62 \\ \hline
\multirow{3}{*}{B97M-rV} & GTH/def2-TZVPPD & 4.22 & 5.96 & 7.85 & 11.38 & 10.19 \\
 & GTH-NLCC-OPT-all$^a$/def2-TZVPPD & 25.61 & 10.26 & 9.03 & 30.51 & 18.92 \\
 & GTH-NLCC-OPT$^b$/def2-TZVPPD & 2.86 & 5.10 & 7.17 & 4.17 & 2.80 \\
 \bottomrule[1.5pt]
\end{tabular}%
\end{center}}
\tabnote{
$^{\rm a}$ GTH-NLCC-OPT-all means that NLCC parameters are optimized for all elements including the second and third rows.\tabnote{$^{\rm b}$ GTH-NLCC-OPT means that NLCC parameters are optimized only for the second-row elements.}

\begin{figure}[H]
\centering
\includegraphics[scale=0.8]{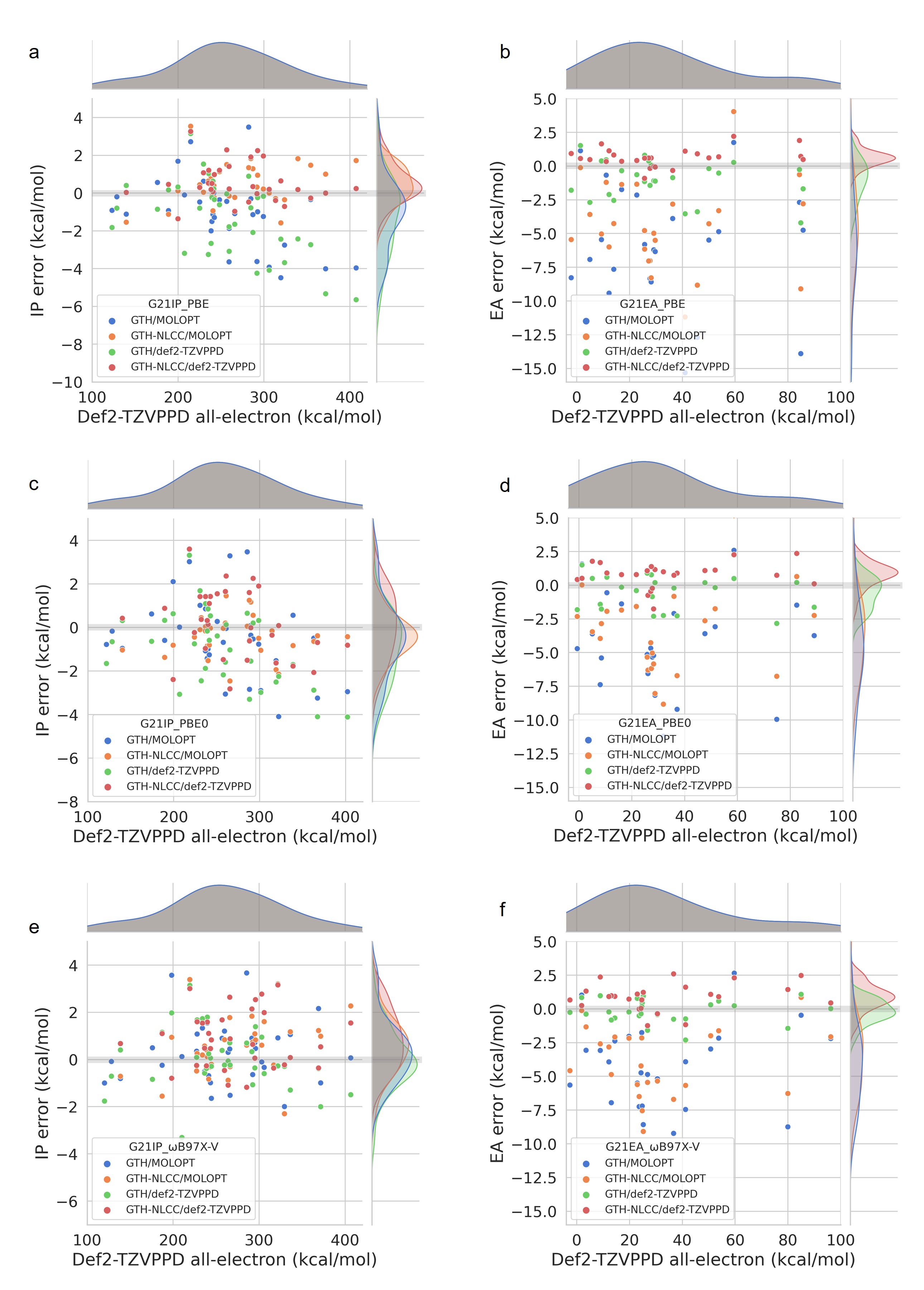}
\caption{Distributions of MADs and all-electron calculations at the levels of PBE, PBE0 and $\omega$B97X-V for IP and EA datasets\cite{doi:10.1063/1.460205,doi:10.1021/ct100466k}.}
\end{figure}

\begin{figure}[H]
\centering
\includegraphics[scale=0.7]{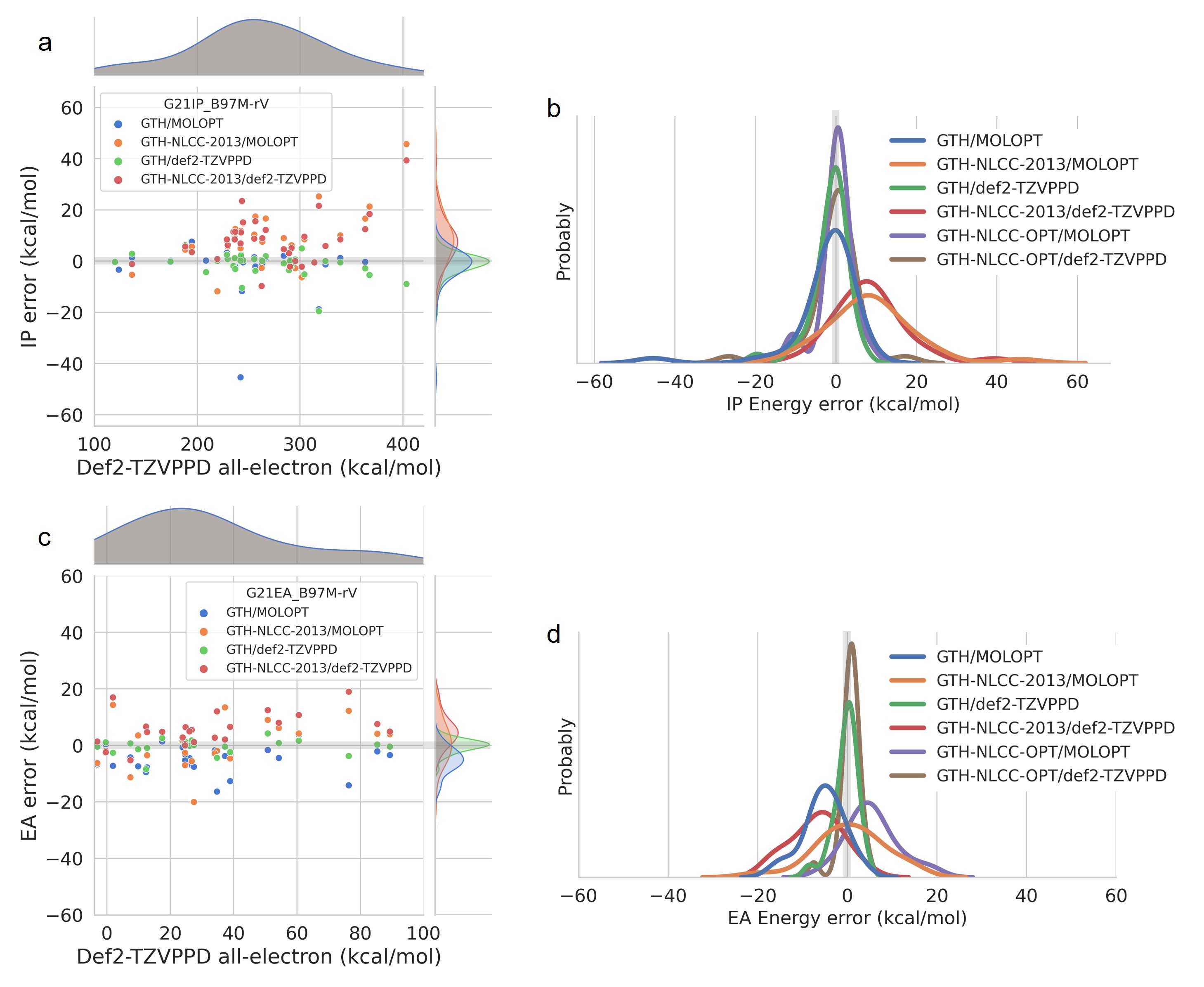}
\caption{Distributions of MADs and all-electron calculations at the level of B97M-rV for IP and EA datasets.}
\end{figure}

\begin{figure}[H]
\centering
\includegraphics[scale=0.75]{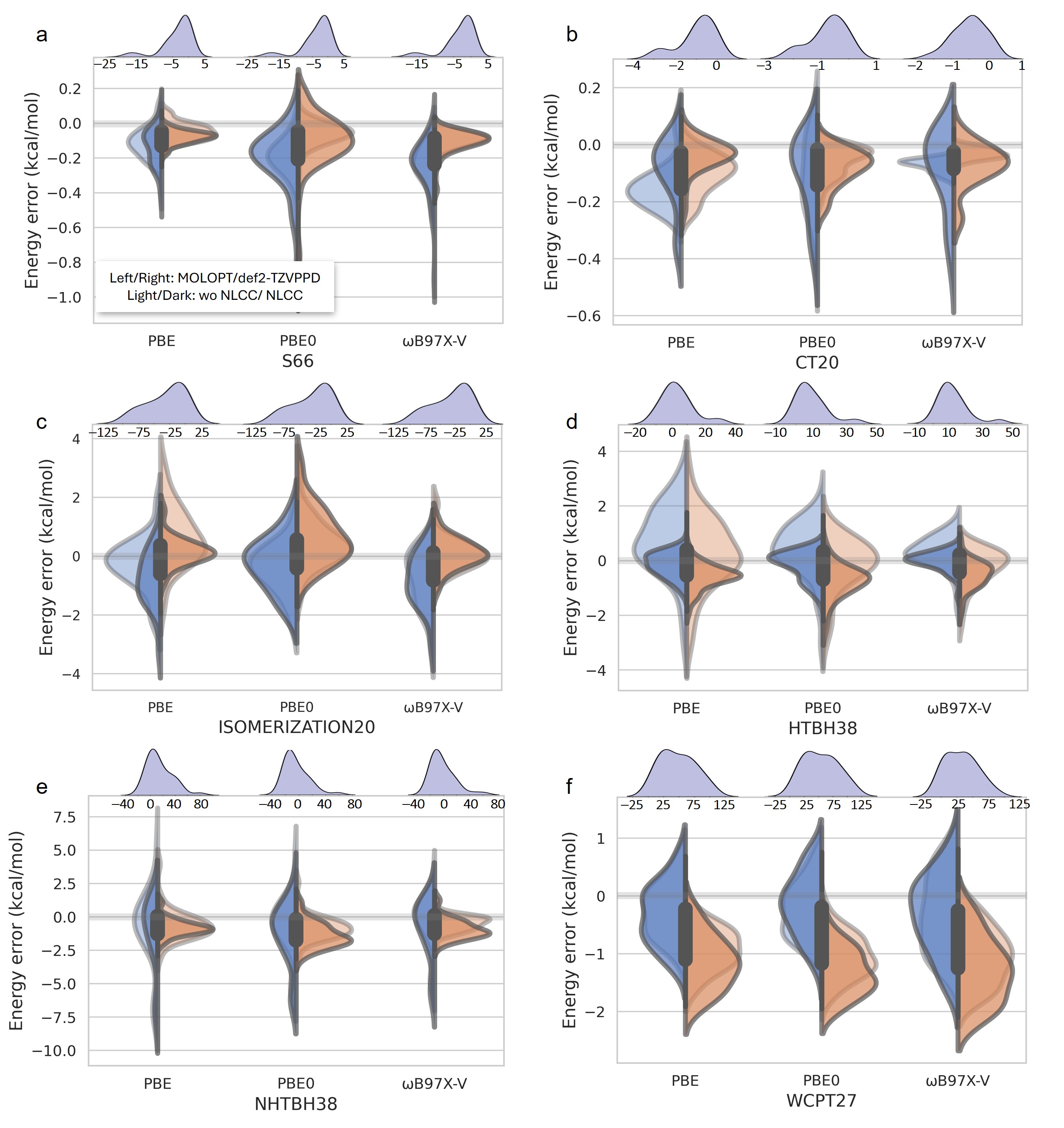}
\caption{NLCC performance on non-thermochemical properties. The energy errors are compared with def2-TZVPPD all-electron calculations, of which the distribution and range are shown by the density plot on the top of each subplot.}
\end{figure}

\bibliographystyle{chem-acs}
\bibliography{PP}